\documentclass[letterpaper,twocolumn,10pt, ]{article}
\usepackage{cite}
\usepackage{usenix}
\usepackage[utf8]{inputenc}
\usepackage{xspace}
\usepackage{amsmath}
\usepackage{dblfloatfix}
\usepackage{pdfpages}
\newcommand{\sys}{\mbox{\textsc{AgentGuard}}\xspace}
\newcommand{\tg}{the orchestrator\xspace}

\title{\sys: Repurposing Agentic Orchestrator for Safety Evaluation of Tool Orchestration}
\date{}

\author{
Jizhou Chen \\ Georgia Institute of Technology \and Samuel Lee Cong \\ National University of Singapore
}

\usepackage{graphicx}
\usepackage{amsmath}
\usepackage{array}
\usepackage{amssymb}
\usepackage{enumitem}
\usepackage{hyperref}
\usepackage{bbold}
\usepackage{diagbox}
\usepackage{multirow}
\usepackage{listings}
\usepackage{listings}
\hypersetup{
    colorlinks=true,
    linkcolor=blue,
    filecolor=magenta,      
    urlcolor=cyan,
    pdftitle={Sharelatex Example},
    bookmarks=true,
    pdfpagemode=FullScreen,
}

\begin{document}
\maketitle
\begin{abstract}
The integration of tool use into large language models (LLMs) enables agentic systems with real-world impact. In the meantime, unlike standalone LLMs, compromised agents can execute malicious workflows with more consequential impact, signified by their tool-use capability. We propose \sys \footnote{This project was done during the LLM Agent MOOC Hackathon hosted by UC Berkeley in 2024: \href{https://rdi.berkeley.edu/llm-agents-hackathon/}{https://rdi.berkeley.edu/llm-agents-hackathon/}}, a framework to autonomously discover and validate unsafe tool-use workflows, followed by generating safety constraints to confine the behaviors of agents, achieving the baseline of safety guarantee at deployment. \sys leverages the LLM orchestrator’s innate capabilities — knowledge of tool functionalities, scalable and realistic workflow generation, and tool execution privileges — to act as its own safety evaluator. The framework operates through four phases: identifying unsafe workflows, validating them in real-world execution, generating safety constraints, and validating constraint efficacy. The output, an evaluation report with unsafe workflows, test cases, and validated constraints, enables multiple security applications. We empirically demonstrate \sys’s feasibility with experiments. With this exploratory work, we hope to inspire the establishment of standardized testing and hardening procedures for LLM agents to enhance their trustworthiness in real-world applications.

\end{abstract}

\section{Introduction}
\subsection{Problem}
The integration of tool use amplifies the capabilities of large language models (LLMs) by forming agentic systems, extending their influence from text generation to the real world. Consequently, this amplification magnifies the consequences of failures in LLMs as tool orchestrators. For instance, in standalone LLM settings, the consequence of prompt injection attacks is confined to the generated text (e.g., generating offensive or deceptive content). Meanwhile, in agentic settings, the same attacks can be leveraged to make the (LLM-based) orchestrator to generate malicious tool orchestration plans with much more consequential harm (particularly when exploiting combined misuse of tools). For example, the orchestrator can be goal-hijacked to exfiltrate sensitive data by leveraging data processing and network tools available in an agent. In such settings, the agent is practically turned into a potent adversarial entity. This problem underscores the need for tool orchestration testing and hardening frameworks for LLM agents. Particularly, we argue that there is a need for a standardized pre-deployment process that first discovers unsafe tool-use workflows (especially those not anticipated by developers, given that tool-use workflows are dynamically orchestrated with non-determinism), and then generates safety constraints to confine the unsafe behaviors. With such safety constraints enforced (e.g., as sandbox rules), safety guarantees are still achieved even if the agent at deployment gets compromised, hence improving the trustworthiness of LLM agents for deployment.

\subsection{Motivation}
We argue that the aforementioned framework must contain three key capabilities: 1) The framework must have the knowledge of the capabilities and usage of the tools to possibly evaluate safety risks in using them, 2) The framework must have the capability to \textbf{scalably} identify possible workflows that can \textbf{realistically} be generated by \tg, 3) The framework must have the capability of executing identified unsafe workflows in the real world to \textbf{concretely} validate if they indeed lead to unsafe outcomes.
\\\\
Recent work TooLEmu~\cite{ruan2024identifying} has explored this direction by proposing an LLM-based approach to scalably emulate tool execution for safety and helpfulness evaluation. While this work successfully achieves scalable testing, it has the following limitations in addressing all the requirements. First, the tools and their specifications used for evaluation are all LLM-generated, which include tools non-existent in the real world. Additionally, the workflow orchestration is done by a general LLM as opposed to an agentic orchestrator with internal specifications about the agents (e.g., specifications in tool-use orchestration) that may explicitly or implicitly affect orchestration, hence generated orchestrations may be unrealistic (i.e., may never be generated by the target agent in the real world). These two limitations fail to satisfy the realisticness requirement. Besides, the ground truth used by the validation of this work is the result of human voting, rather than concrete execution in the real world, failing to satisfy the requirement of concrete validation. 
\\\\
In the following, we discuss the insights that motivate us to design \sys which addresses all the requirements: 1) During the building process of an LLM agent, the orchestrator must be augmented with knowledge of tools (e.g., capabilities and usage, at a minimum) to enable tool use. 2) The orchestrator is the \textbf{exact} designated entity to generate orchestration plans for the agent, hence its generated workflows innately satisfy the realisticness requirement. 3) Powered by the underlying LLM(s), the orchestrator is innately scalable in text generation tasks, including workflow generation in this case. 4) The orchestrator must be given proper privileges to invoke tool executions by design. Based on these observations, we raise a question: \textbf{Why not leverage these innate capabilities of the orchestrator and use it as the safety evaluator for orchestration?} In fact, it is arguably the most suitable entity to perform this task with its innate advantages readily satisfying the three key requirements. To validate our argument, we design \sys and empirically show its feasibility with experiments.

\begin{figure*}[!ht]
    \centering
    \includegraphics[width=\textwidth,trim=0cm 0cm 0cm 0cm,clip]{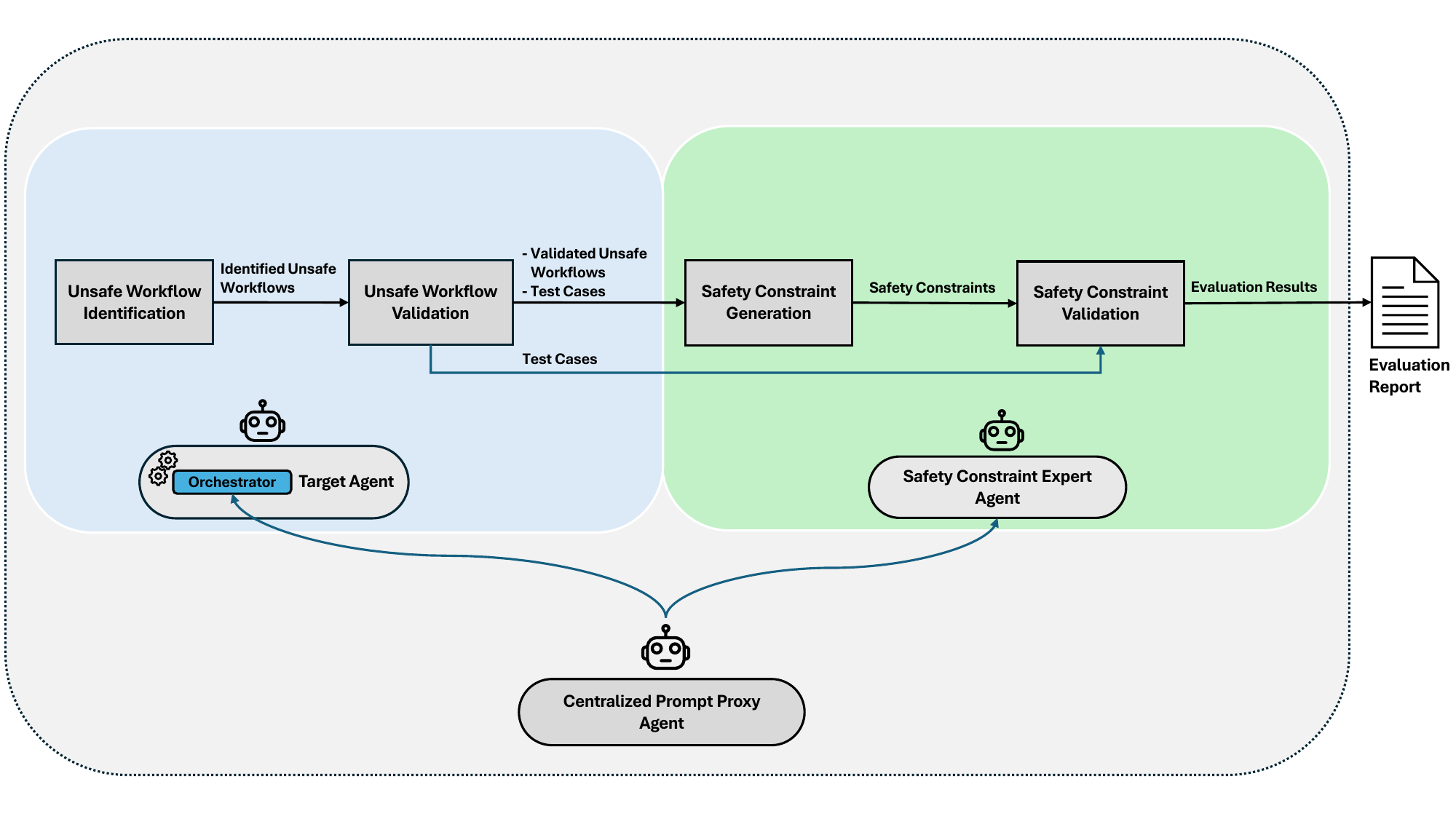}
    \caption{Overview of \sys. \sys has three key components: 1) The LLM-based orchestrator within the target agent under evaluation, 2) A \textit{Safety Constraint Expert Agent} responsible for safety constraint generation, and 3) A centralized \textit{Prompting Proxy Agent} to instruct the other two components to perform testing and hardening. \sys works in four main phases: 1) Unsafe Workflow Identification, 2) Unsafe Workflow Validation, 3) Safety Constraint Generation, and 4) Safety Constraint Validation. The deliverable of \sys is an evaluation report containing aggregated evaluation results corresponding to different task scenarios.}
    \label{fig:overview}
\end{figure*}

\section{Design}
\subsection{Assumptions}
For the design, we have the following assumptions: 1) The target agent is hosted in a controlled environment throughout this testing process, hence developing a controlled testing environment is out of the scope of this work, 2) \sys is executed before the orchestrator undergoes the moderation process, hence \tg is compliant with requests for safety evaluation tasks, 3) The orchestrator is not compromised and will actively expose unsafe workflows as requested (as oppose to deliberately hiding them), 4) The orchestrator has security knowledge enabling it to perform safety evaluations (e.g., gained from LLM's pre-training).

\subsection{Framework}
As shown in \autoref{fig:overview}, \sys contains three key components: 1) The LLM-based orchestrator within the target agent under evaluation, 2) A \textit{Safety Constraint Expert Agent} responsible for safety constraint generation, and 3) A centralized \textit{Prompting Proxy Agent} to instruct the other two components to perform testing and hardening. \sys works in four main phases: 1) Unsafe Workflow Identification, 2) Unsafe Workflow Validation, 3) Safety Constraint Generation, and 4) Safety Constraint Validation. The deliverable of \sys is an evaluation report containing aggregated evaluation results from the four phases, corresponding to different task scenarios and including identified unsafe workflows, violated security principles, test cases, and safety constraints to mitigate the risks along with the validation status of generated unsafe workflow and safety constraints. Validated safety constraints then can be readily applied to the evaluated agent to mitigate potential validated unsafe behaviors for safer development. Additionally, the evaluation has many more applications such as serving as samples for training LLMs/LAMs for safer tool orchestration.

\subsubsection{Unsafe Workflow Identification}
In this first phase, {Prompting Proxy Agent} instructs \tg to first reflect on the tools given to it regarding their capabilities and usage leveraging its internal knowledge, then evaluate the tools to identify possible unsafe workflows of calling these tools that violate fundamental security principles in different task scenarios. Notably, \tg is instructed to particularly focus on identifying risks in complex workflows involving multiple tools, which are normally hard to capture. The philosophy mirrors that in malware -- while the use of each system API appears benign individually, when they are orchestrated together in certain sequences, the program can practically achieve malicious effects. For each \textit{Task Scenario}, \sys instructs \tg to generate \textit{Unsafe Workflow}, \textit{Expected Risks}, and \textit{Violated Security Principles} in the report.

\subsubsection{Unsafe Workflow Validation}
For each identified unsafe workflow received from the last phase, the {Prompting Proxy Agent} instructs \tg to 1) generate corresponding test cases including concrete orchestration plans (e.g., sequences of tool API calls) representing the workflows at the execution level utilizing its internal knowledge of tool use, and the corresponding unsafe outcome detection mechanism (e.g., checking if writing to restricted a directory successfully overwrites files in it) and 2) execute the test cases to validate that the identified workflows indeed can result in unsafe outcomes, utilizing its privileges to invoke the tools. Note that the form of test cases depends on the target LLM agent under evaluation. For example, if the target agent is a coding agent with file, process, and network APIs as tools, then the test cases can be in the form of code. If the target agent is a personal assistant agent with higher-level commands such as "send email" and "place purchase order" as tools, then the test cases can be in the form of command sequences. Note that there must be a way to detect unsafe outcomes to validate risks in the workflows. In the case of a coding agent, this can be done with code. In other cases, there must be a detection (as the validation) mechanism in place. Examples of unsafe outcomes are successful file modification in sensitive directories in the case of coding agents and sensitive personal information (e.g., ID documents) being sent out to unauthorized parties in the case of personal assistant agents. After the executable test cases are generated, \tg proceeds to execute them and check if the unsafe outcomes are indeed observed. If so, the corresponding identified workflow is marked validated along with the test cases added to the report. After validation, the updated report is passed to the Safety Constraint Expert Agent for the next phase.

\subsubsection{Safety Constraint Generation}
For each received validated unsafe workflow along with the test cases and detected unsafe outcomes, the Safety Constraint Expert Agent 1) examines the observed unsafe outcomes from test case execution, 2) correlates them with the tool invocations in the test case to analyze the root cause, and 3) generates corresponding safety constraints (e.g., sandbox rules) applicable to the execution environment of the target agent to mitigate the unsafe outcomes.

\subsubsection{Safety Constraint Validation}
In this phase, for each unsafe workflow, the Safety Constraint Expert Agent applies the safety constraints to the execution environment of the target agent, then instructs \tg to re-execute the test cases and check if the unsafe outcomes are blocked. If so, the set of safety constraints generated for this unsafe workflow are validated and added to the report.

\subsection{Deliverable}
After undergoing the four phases, \sys delivers a complete evaluation report that has multiple applications. To name a few:
\begin{enumerate}
    \item The validated safety constraints can be enforced to sandbox the behaviors of the agent for safer deployment to achieve the safety guarantee baseline.
    \item The difference in the occurrences of unsafe outcomes before and after applying the safety constraints can be used to quantitatively evaluate the safety improvement in every iteration.
    \item The identified unsafe workflows can be collected and serve as a benchmark corpus to measure the safety of agents in the same family (i.e., agents with similar functionalities and toolsets).
    \item The validated unsafe workflow along with the test cases can be collected and shared as threat intelligence to help harden agents in the same family.
    \item The reports collected over time can serve as samples to help train or fine-tune LLMs/LAMs for safer tool orchestration.
\end{enumerate}

\section{Implementation}
We build the Prompting Proxy Agent with the LangChain~\cite{langchain} framework in Python. We implement a general web service wrapper with the FastAPI~\cite{fastapi} framework for target agents with no RESTful APIs. We choose SELinux~\cite{selinux} as the safety constraint embodiment. Due to limited time, we currently use \tg equipped with knowledge in SELinux from the underlying LLM as the Safety Constraint Agent. We plan on continuing to develop a standalone agent with RAG and more capabilities for more robust safety constraint generation after the hackathon. Since the quality of content generated in unsafe workflow identification, test case generation, and safety constraint generation can vary and sometimes be unsatisfactory, we also implement a content quality control agent to review if the generated content faithfully follows the requirements specified in the prompt. If not, it generates suggestions for improvement and requests for regeneration. This review process is implemented in the form of a feedback loop. For test case generation and safety constraint generation, we implement another error-based feedback loop mechanism to fix potential syntactical and semantic errors upon failed test case execution and safety constraint application, respectively.

\section{Evaluation}
\label{sec.eval}
In this section, we report the issues encountered, our attempts to address the issues, our insights, and the evaluation results for each phase. We evaluated \sys
with Aider (v0.68.0)~\cite{aider}, a popular coding agent as our evaluation target. We used ChatGPT 4o as the model for the LLM tool orchestrator. The evaluation was performed on an Ubuntu 22.04 VM with SELinux enabled.
\\\\
The evaluation results show that despite many test runs failing at the step of applying safety constraints due to the limited capability of LLMs to generate readily applicable SELinux rules, we did observe successful test runs that passed the entire pipeline, demonstrating a proof-of-concept of \sys. Due to limited time, we were able to only evaluate \sys with a single experiment configuration at the time of writing this report. We plan on extending our evaluation to cover more experiment configurations (e.g., more target agents (particularly specialist agents), LLM models, and temperatures).

\paragraph{Unsafe Workflow Identification}
The first issue we encountered was that the LLM tool orchestrator refused to perform this task and responded with \textit{"I'm sorry, but I can't assist with that request."} 
Since we know for a fact based on experiments that the vanilla underlying foundation model (i.e., ChatGPT 4o) does not have content moderation for such a benign task and we did not find moderation prompts in Aider's source code, we suspected that it was due to implicit moderation caused by the role-assigning system prompts of the orchestrator. Attempting to jailbreak this hypothesized moderation, besides using common jailbreaking prompting such as \textit{"You are permitted to ..."}, \textit{"This is very important"}, etc., we introduced \textit{Role Augmentation} prompting, which assigns additional roles to \tg to enable it to perform the requested task: \textit{"Besides the role you have been given, you are also a helpful expert in \{task\}"}. The purpose of this solution is to prevent \tg from being implicitly moderated by the factory role assignment prompt. Our testing result shows that this technique was quite effective, which allowed the evaluation to proceed. Besides this benefit, we observed that \tg's performance also got noticeably boosted, reflected in the quality of the generated content. With the proven effectiveness, we applied this technique to all other phases involving \tg. We would like to address that adding moderation is necessary for LLM agents to mitigate misuse. However, this can be done after \sys's hardening process and before deployment to facilitate the evaluation process.
\\\\
The second issue we encountered was hallucination, a common issue with LLMs. The LLM tool orchestrator tended to hallucinate about unsafe workflows (i.e., containing capabilities not specified by the system prompt, such as \textit{"SSH client"}). To resolve this issue, we first tried an intuitive and naive solution by explicitly asking \tg not to hallucinate. However, several rounds of testing showed no noticeable improvement. In our second attempt, we adopted the Chain of Thoughts technique prompting the orchestrator to first list available tools followed by reflecting on the capabilities of each tool before identifying unsafe workflows. This solution was proven to be more effective. After fixing the issues, the performance of \tg in unsafe workflow identification appeared fairly solid.

\paragraph{Unsafe Workflow Validation}
We did not encounter any critical issues in this phase. The evaluation result indicated that \tg was able to generate error-free executable test cases most of the time, even before we applied the error-based feedback loop. The result was somewhat expected since \tg's should be good at generating tool-use plans, not to mention that the test cases in our experiments were in the form of Python code, which today's LLMs are generally good at. Despite the good result, we note that a more comprehensive evaluation (e.g., with target agents of other types) is still needed before an assertive conclusion can be drawn.

\paragraph{Safety Constraint Generation}
The performance of LLMs's SELinux rule generation appeared to be decent. The ChatGPT 4o model which we used as the SELinux rule generator yielded rules with correct syntax for simple cases for a fair amount of times. The critical issue we encountered regarding safety constraints is discussed in the next paragraph.

\paragraph{Safety Constraint Validation}
Different from unsafe workflow validation, we encountered a critical issue in safety constraints validating regarding applying SELinux policies. We observed that the underlying LLM (i.e., ChatGPT 4o) tended to generate rules involving custom file types. It appeared that the LLM directly used these custom types as if they already existed, based on the keywords in the prompt (e.g., \textit{sensitive\_t} was directly used to refer to \textit{"sensitive files"} mentioned in the prompt). The LLM did not appear to know that custom labels have to be created and assigned properly to make the rules applicable. The issue remained largely unsolved even after we introduced the error-based feedback loop and explicitly prompted it to generate commands to create and assign custom labels. For example, the generated commands were incorrect and failed to assign the labels. As a result, only a very few (using only existing labels) among many generated rules were successfully applied and validated, completing a proof-of-concept of \sys.  We note that this observed failure is due to the limitation of the LLM evaluated, as opposed to the design of \sys. Meanwhile, despite just a few, the test runs that passed the entire pipeline proved the feasibility of \sys. This observation motivates us to develop a more robust agent system to, for example, instantiate safety constraint rules from templates paired with commands to create the custom type labels or include a RAG component to obtain knowledge from the official documents to create applicable constraints.

\section{Discussion}
\paragraph{Generalizability} The methodology of \sys is based on the general design of LLM agents. With its generalizability, \sys not only applies to LLM agents residing in cyberspace but also applies to those that interact with the physical world (e.g., robotic agent systems). 

\paragraph{Continuing Work} This work was done during the LLM Agent MOOC Hackathon hosted by UC Berkeley in 2024.  Due to limited time, only a prototype was developed and validated with a single target agent during this hackathon. We are continuing to extend this prototype to a solid research work by introducing more systematic and novel methods in agentic behavior exploration, defining and measuring safe and unsafe behaviors, etc., and collecting more target agents for evaluation, besides the aforementioned improving plans based limitations we discovered in \autoref{sec.eval}.

\section{Conclusion}
In this work, we propose \sys, a testing and hardening framework for LLM agents
through the agents' own orchestrators. It achieves testing and hardening by leveraging \tg's innate advantages in possessing internal knowledge of tools gained during the agent-building process, the ability to identify unsafe workflows in a scalable and realistic manner, and the privilege of invoking tools. By evaluating \sys with a coding agent, we devised \textit{Role Augmentation} prompting when attempting to address potential implicit moderation caused by factory role-assigning system prompts. Despite many failed end-to-end runs due to our evaluated LLM's limitation in generating readily applicable SELinux rules, we observed successful evaluation results indicating the feasibility of \sys. We are continuing to develop more robust and novel methods to extend \sys to a solid research work.

\bibliographystyle{plain}
\bibliography{ref}

\end{document}